\documentclass[]{spie}  

\usepackage{amsmath,amsfonts,amssymb}
\usepackage{graphicx}
\usepackage[colorlinks=true, allcolors=blue]{hyperref}
\usepackage{siunitx}
\usepackage{caption}
\usepackage{subcaption}

\title{A Polarization Modulator Unit for the Mid- and High-Frequency Telescopes of the LiteBIRD mission}

\author[a, b]{Fabio Columbro}
\author[a, b]{Paolo de Bernardis}
\author[a, b]{Luca Lamagna}
\author[a, b]{Silvia Masi}
\author[a, b]{Alessandro Paiella}
\author[a, b]{Francesco Piacentini}
\author[d]{Giampaolo Pisano\textsuperscript{a,c}, for the Litebird Joint Study Group}

\affil[a]{Dipartimento di Fisica, Sapienza Universit\`a di Roma, P.le A. Moro 5, 00185, Roma, Italy}
\affil[b]{INFN--Sezione di Roma1, P.le A. Moro 5, 00185, Roma, Italy} 
\affil[c]{School of Physics and Astronomy, Cardiff University, Queens Buildings, The Parade, Cardiff, CF24 3AA, U.K.} 
\affil[d]{LiteBIRD Joint Study Group Members are listed at \href{https://wiki.kek.jp/display/cmb/LiteBIRD+Joint+Study+Group+members+picture+book}{this link}}

\authorinfo{Further author information: (Send correspondence to Fabio Columbro)\\ E-mail: fabio.columbro@roma1.infn.it}

\begin{document} 
\maketitle

\begin{abstract}
The LiteBIRD mission is a JAXA strategic L-class mission for all sky CMB surveys which will be launched in the 2020s. The main target of the mission is the detection of primordial gravitational waves with a sensitivity of the tensor-to-scalar ratio $\delta r<0.001$. The polarization modulator unit (PMU) represents a critical and powerful component to suppress $1/f$  noise contribution and mitigate systematic uncertainties induced by detector gain drift, both for the high-frequency telescope (HFT) and for the mid-frequency telescope (MFT).
Each PMU is based on a continuously-rotating transmissive half-wave plate (HWP) held by a superconducting magnetic bearing in a \SI{5}{\kelvin} environment.
In this proceeding we will present the design and expected performance of the LiteBIRD PMUs and testing performed on every PMU subsystem with a room-temperature rotating disk used as a stand-in for the cryogenic HWP rotor.
\end{abstract}


\section{INTRODUCTION}
\label{sec:intro}  

The \textit{Lite (Light) satellite for the studies of B-mode polarization and Inflation from cosmic background Radiation Detection} (LiteBIRD) mission\cite{Sugai:article} is the successor of the CMB space missions COBE\cite{COBE:ARTICLE}, WMAP\cite{WMAP:article}, and Planck\cite{Planck:article}, each of which has given landmark scientific discoveries. A detection of primordial gravitational waves with LiteBIRD (at a level $\delta r < 0.001$) would indicate that inflation occurred near the energy scale associated with grand unified theories and would provide additional evidence of an inherently quantum-gravitational process\cite{Grav_waves}.
Additionally, the energy scale of inflation has important implications for other aspects of fundamental physics, such as axions and neutrinos. LiteBIRD’s ability to measure the entire sky at the largest angular scales with 15 frequency bands is complementary to ground-based experiments\cite{LiteBIRD_PTEP}. Ground-based experiments can also improve LiteBIRD observations with high-resolution lensing data.

A key component of LiteBIRD is its polarization modulator unit (PMU), an essential feature to suppress the $1/f$ noise contribution (low-frequency system drifts induced by thermal variations or detector gain drift) and and mitigate systematics uncertainties induced by detector gain drifts.
The polarization modulation methodology based on HWP is already used by a large number of experiments and can be divided in two families: a step-and-integrate strategy (SPIDER\cite{SPIDER:article}, QUBIC\cite{QUBIC_hwp, QUBIC:article}) and a continuously-rotating HWP (ABS\cite{ABS:article}, EBEX\cite{ebex_pol}, ACT-pol\cite{ACTpol:article}, POLARBEAR-2\cite{Polarbear:article, Polarbear2}, LSPE/SWIPE\cite{Lamagna:article, Columbro2020:article}).
Each of the 3 telescopes (high, mid and low frequency: HFT, MFT, LFT) is equipped with a cryogenic continuously-rotating half-wave plate (HWP) based on a superconducting magnetic bearing (SMB), an emerging technology with a low technology readiness level (TRL).

In this contribution we present the baseline design  (Sec.~\ref{sec:baseline}) of the MHFT (mid- and high-frequency telescope) PMUs, which use the metal-mesh filter technology, while the LFT PMU uses of an achromatic 9-layer sapphire HWP\cite{Sakurai_SPIE}. The MHFT design takes inspiration from the LSPE/SWIPE one which is under testing (Sec.~\ref{sec:mockup}). A room-temperature mockup was used to develop and validate the eddy current model and to develop the driver and readout electronics (Sec.~\ref{sec:electronic}).
The expected performance of the LiteBIRD PMUs is discussed in Sec.~\ref{sec:performance} and will be confirmed during the first test of the breadboard model.

\section{Baseline design}
\label{sec:baseline}
In this section, we present the baseline design of the LiteBIRD MHFT PMU. Since both modulators will be mounted on a space mission, the design is driven by stringent requirements on mass, dimensions, stiffness, power dissipation, and TRL for the levitation, driving, gripping, and position encoding mechanisms. The most important requirements for both PMUs are summarized in Tab.~\ref{tab:MHFT_requirements}.

\begin{table}[ht]
\caption{MHFT-PMU main requirements.}
\label{tab:MHFT_requirements}
\begin{center}       
\begin{tabular}{|l|l|l|} 
\hline
\rule[-1ex]{0pt}{3.5ex}  \bf{Parameter} & \multicolumn{2}{c|}{\bf{Requirement}}  \\
\hline
\rule[-1ex]{0pt}{3.5ex}   & \bf{MFT} & \bf{HFT}   \\
\hline
\rule[-1ex]{0pt}{3.5ex}  Spin rate & \SI{39}{rpm} (\SI{0.65}{\hertz}) & \SI{61}{rpm} (\SI{1.02}{\hertz})  \\
\hline
\rule[-1ex]{0pt}{3.5ex}  HWP diameter & \SI{320}{\milli\meter} & \SI{220}{\milli\meter}  \\
\hline
\rule[-1ex]{0pt}{3.5ex}  HWP temperature & \multicolumn{2}{c|}{$<\SI{20}{\kelvin}$}  \\
\hline 
\rule[-1ex]{0pt}{3.5ex}  Load on the \SI{5}{\kelvin} stage & \multicolumn{2}{c|}{$<\SI{4}{\milli\watt}$}  \\
\hline 
\rule[-1ex]{0pt}{3.5ex}  Angular accuracy & $<\SI{1}{\arcmin}$ & $<\SI{5}{\arcmin}$ \\
\hline 
\rule[-1ex]{0pt}{3.5ex}  Total mass & \multicolumn{2}{c|}{$<\SI{20}{\kilogram}$}  \\
\hline 
\end{tabular}
\end{center}
\end{table}

\begin{figure} [ht]
   \begin{center}
   \begin{tabular}{c} 
   \includegraphics[height=7cm]{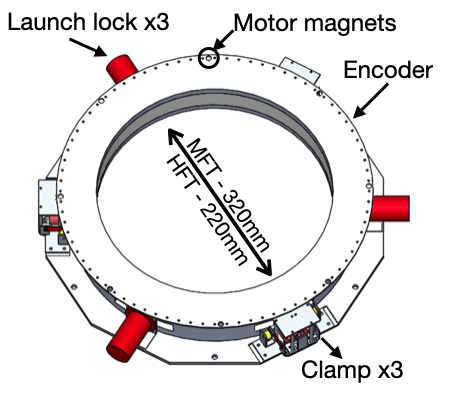}
\includegraphics[height=7cm]{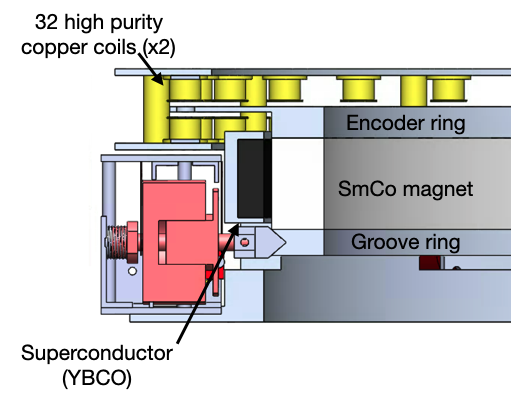}
   \end{tabular}
   \end{center}
   \caption[example] 
   { \label{fig:MHFT_design}
     \textit{Left panel}: Overview of the polarization modulator unit design. The concept of the design is the same with a scaling of the geometry (\SI{320}{\milli\meter} and \SI{220}{\milli\meter} diameter HWPs are mounted in the center, for MFT and HFT, respectively). \textit{Right panel}: Section view of the modulator: a nearly frictionless  bearing is obtained with the magnetic levitation of rotor composed by a permanent magnet rotor ring (cream grey) sandwitched between an encoder ring and a groove ring. The stator ring hosts an array of superconducting bulks (black) and the electromagnetic motor composed of 2 sets of 32 coils each coupled with 8 small motor magnets hosted in the rotor.
   }
\end{figure}

The modulator is conceptually similar to the EBEX\cite{SMB:article, ebex_pol}, POLARBEAR-2\cite{Polarbear2} and LSPE/SWIPE\cite{LSPE:article} designs, but is more challenging and ambitious because of the space application.
The HWP diameters of MFT and HFT are \SI{320}{\milli\meter} and \SI{220}{\milli\meter}, respectively.
The concept of the design is shown in Fig.~\ref{fig:MHFT_design} and is the same for both modulators with a scaling of the components.

In contrast to the most common design of a superconducting magnetic bearing (SMB), we chose a different configuration: the magnet ring and the superconductor are not stacked up but the internal rotating ring is the magnetic one and the external is the superconducting one, in order to obtain a side face to face interaction and minimize horizontal displacement.

The selected superconductor, YBCO (Yttrium barium copper oxide), is the type-II superconductor with the highest
pinning force and critical current density ($\sim$ \SI{e5}{\ampere.\milli\meter^{-2}}), and was chosen because higher critical current means lower hysteresis losses\cite{Bean:article}.
The rotor is composed of three rings stacked along the optical axis, starting from the bottom:
\begin{itemize}
    \item Groove ring: used to clamp the plate above the YBCO transition temperature.
    \item Magnetic ring: composed of 2 Samarium-Cobalt magnetic rings sandwiched between 3 thin iron rings to produce a more uniform magnetic field.
    \item Aluminum ring with three different functions: to align the HWP in the center, to measure the angular position of the rotor with the encoder and to hold the motor magnets used in the driver system.
\end{itemize}

The drive mechanism is conceptually similar to an electromagnetic motor. We use 8 SmCo magnets (\SI{2}{\milli\meter} thick, \SI{9}{\milli\meter} diameter) coupled with 2 rings of 32 coils each, on the top and on the bottom of the rotor to obtain a larger and more uniform force. The coils are connected in series (4 series of 16 coils each). The geometric parameters chosen are reported in Tab.~\ref{tab:MHFT_coils}, and the average force produced by the motor during operation (16 coils) is \SI{280}{\milli\newton\per\ampere}/\SI{414}{\milli\newton\per\ampere} for MFT/HFT.

\begin{table}[ht]
\caption{Coil parameters for the MFT and HFT. The diameter of the copper wire is \SI{0.2}{\milli\meter} and the resistance reported is assumed at \SI{300}{\kelvin}. }
\label{tab:MHFT_coils}
\begin{center}       
\begin{tabular}{|l|l|l|l|} 
\hline
\rule[-1ex]{0pt}{3.5ex}  \bf{Parameter} & \bf{Unit} & \bf{HFT} & \bf{MFT}   \\
\hline
\rule[-1ex]{0pt}{3.5ex}  \bf{Coil diameter} & \SI{}{\milli\meter} & \SI{6}{} & \SI{5}{}  \\
\hline
\rule[-1ex]{0pt}{3.5ex}  \bf{Coil length} & \SI{}{\milli\meter} & \SI{10}{} & \SI{10}{}  \\
\hline
\rule[-1ex]{0pt}{3.5ex}  \bf{Turn density} & \SI{}{\milli\meter^{-1}} & \SI{25}{} & \SI{25}{}  \\
\hline 
\rule[-1ex]{0pt}{3.5ex}  \bf{Resistance (16 coils)} & \SI{}{\ohm} & \SI{103}{} & \SI{87}{}  \\
\hline 
\end{tabular}
\end{center}
\end{table}

During the launch, the rotor is held above the stator at room temperature by 3 pin pullers\footnote{\url{https://www.ebad.com/tini-pin-puller/}}, radially oriented towards the center of the HWP ring. After the launch, the pin pullers are released and 3 linear actuators hold the rotor in position during the cooldown process, until the YBCO is superconducting and the magnetic field is frozen. Hereafter the rotor is kept in place by the flux pinning and the clamps are released.

We developed a frictionless electromagnetic clamp/release system \cite{Actuator:article} suitable for any experiment equipped with a large cryogenic HWP rotator based on a SMB.
The main features of this system are:
\begin{itemize}
\itemsep-0.3em
\item large rotor mass compliance ($\sim\SI{10}{\kilogram}$);
\item zero power dissipation while holding the rotor;
\item fast ($\sim\SI{40}{\milli\second}$) release with low power dissipation ($\sim\SI{30}{\joule}$) during each operation;
\item low cost and high reliability over hundreds of operation cycles.
\end{itemize}
This system is intended to be used only once but if needed it can clamp the rotor as needed throughout the flight.

The PMU is also equipped with a custom capacitive sensors to measure the temperature and levitation height of the rotor\cite{PdB_levitation_measurement2020}.
The temperature sensor is a thermistor, physically mounted on the rotating device and biased with an AC current, which is transferred from the stationary electronics to the rotating device via capacitive coupling. The levitation height sensor is a network of capacitors, similar to the one used for the capacitive coupling of the thermistor.  The system reaches an accuracy better than 3\% for the measurement of the thermistor resistance, and an accuracy of $\sim\SI{10}{\micro\meter}$ for the measurement of its levitation height.

\subsection{HWP design}
\begin{figure}[h]
   \begin{center}
   \begin{tabular}{c} 
\includegraphics[width=0.48\linewidth]{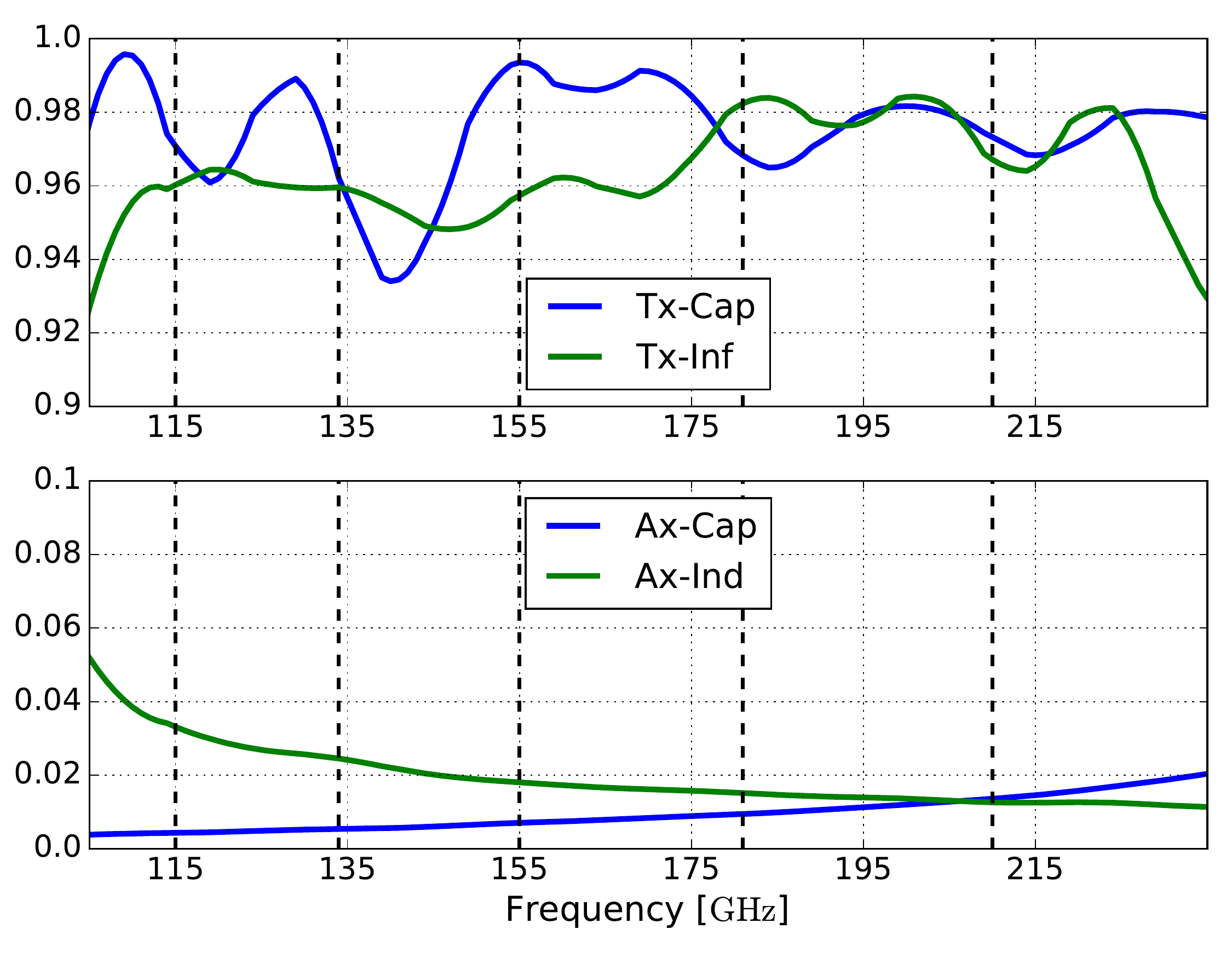}
\includegraphics[width=0.48\linewidth]{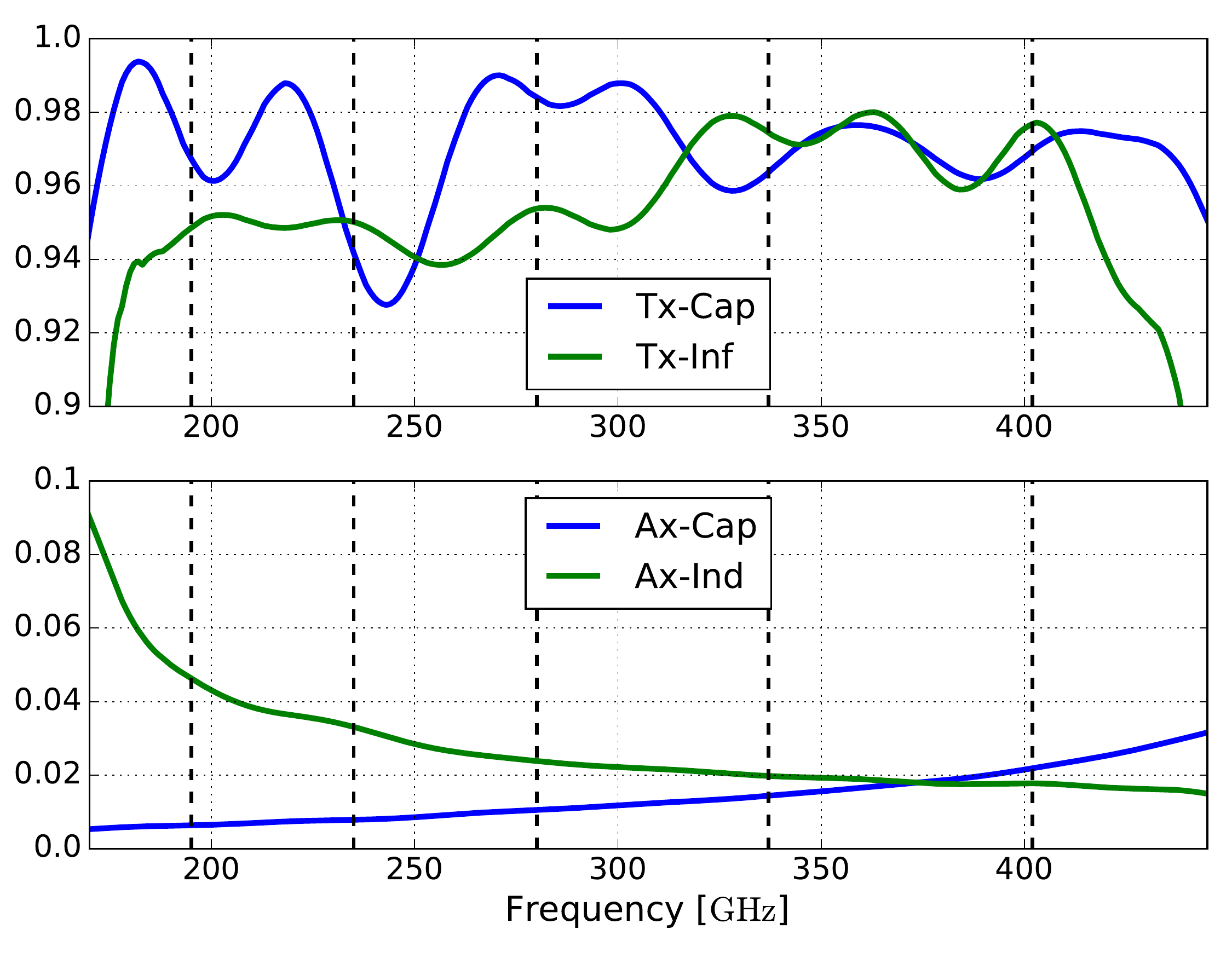}
   \end{tabular}
      \end{center}
\caption{\label{fig:hwp_band}
MFT (\textit{Left}) and HFT (\textit{Right}) mesh-HWP preliminary designs expected performances as a function of frequency: transmissions, absorptions for the capacitive and inductive axes. Vertical dashed lines represent the central frequency of MHFT bands.
}

\end{figure}

The baseline designs for the MFT and HFT HWPs are mesh-HWPs\cite{Pisano1 , Pisano2}. These quasi-optical components are based on the mesh-filter technology\cite{Pisano3}, which has been adapted to mimic anisotropic behaviour. A mesh-HWP is based on two stacks of anisotropic metal grids embedded into polypropylene. The two stacks, one inductive and one capacitive, are designed in such a way that two electromagnetic waves passing through them, polarised in orthogonal directions, will experience \SI{180}{\degree} phase-shift. Each stack has different grids, designed with specific geometries, located at optimised distances. The combination of  all the grids, in our case 5 capacitive and 5 inductive, provides a differential phase-shift around \SI{180}{\degree} across the frequency of operation. The design and the manufacture of mesh-HWPs are described in detail elsewhere\cite{Pisano4}. The expected performance of the MFT and HFT mesh-HWPs are reported in Fig.~\ref{fig:hwp_band}. The transmission coefficients and the modulation efficiencies across the MFT and HFT bands are on average greater than 95\%.

\section{Room-temperature mockup}
\label{sec:mockup}
We developed a room-temperature mockup to validate the motor, the driver and readout electronics, the eddy current model, the main magnet inhomogeneities and the spinning frequency stablity. 
The size of the mockup is similar to the LSPE/SWIPE polarization modulator (\SI{500}{\milli\meter} diameter), but the performance in terms of the friction can be scaled because of the well known diameter dependence (see Sec.~\ref{sec:losses}).
In place of the superconducting magnetic bearing, we used a low-friction ball bearing\footnote{\url{https://www.skf.com/it/index.html}}.

\begin{figure}[h]
   \begin{center}
   \begin{tabular}{c} 
\includegraphics[width=0.94\linewidth]{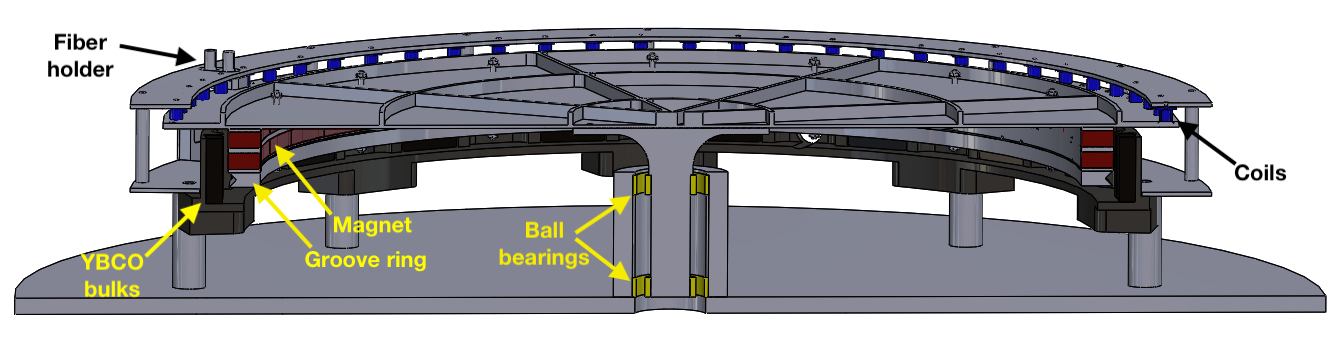} \\
\includegraphics[width=0.65\linewidth]{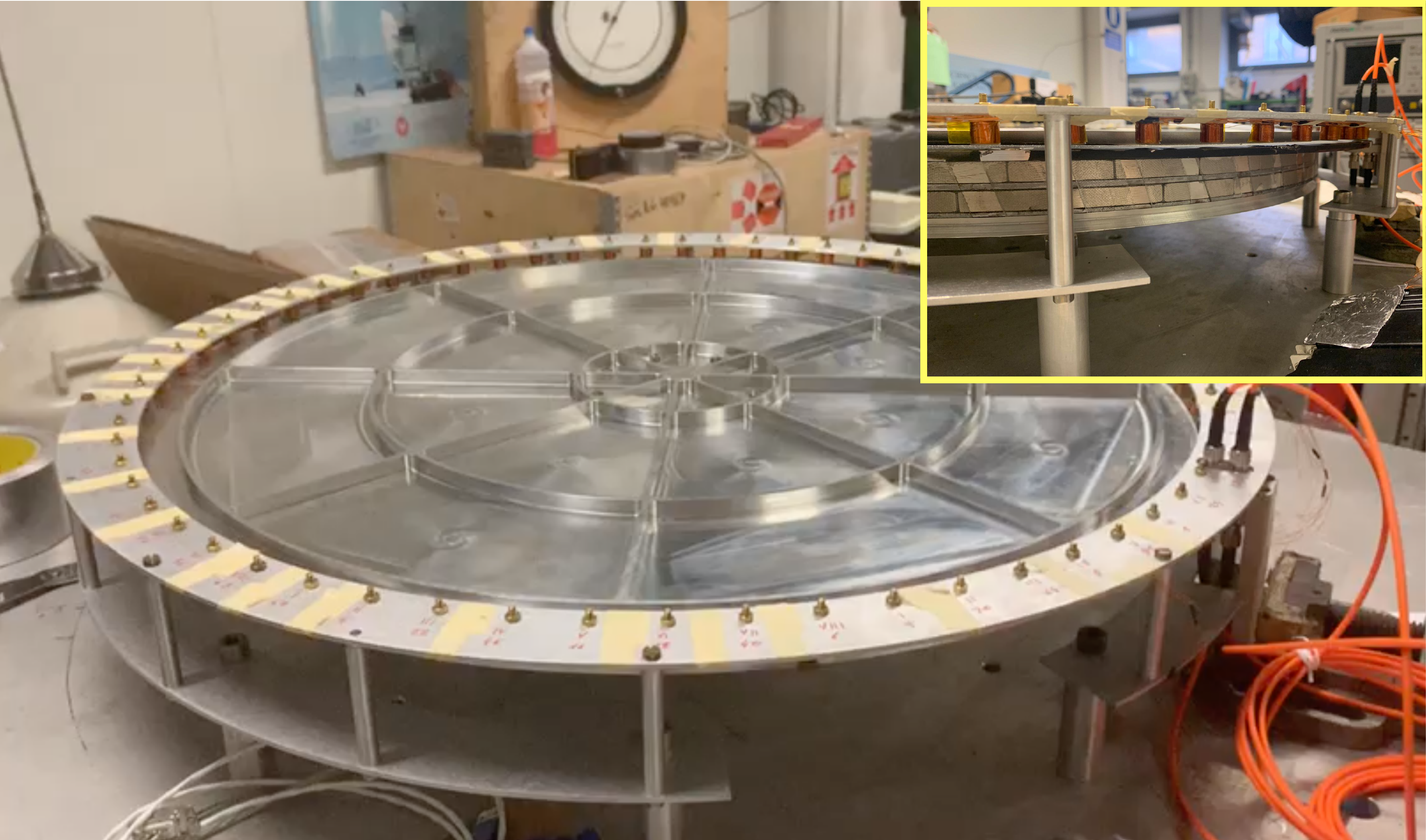}
\includegraphics[width=0.24\linewidth]{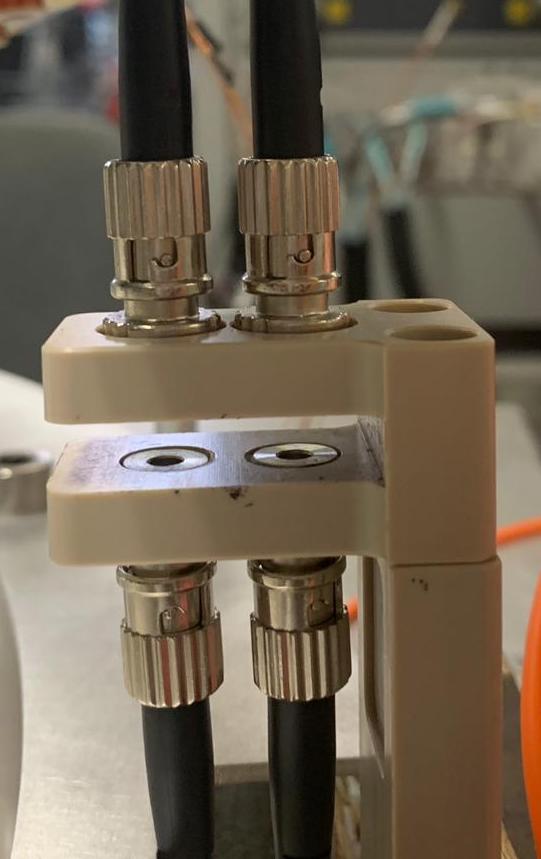}
   \end{tabular}
      \end{center}
\caption{\label{fig:mockup}
(\textit{Top})  CAD section of the room-temperature mockup, composed by 2 ball bearings (yellow) separated by an aluminum spacer allowing to rotate an umbrella support which keeps in position a lightened aluminum disk ($\sim\SI{2}{\kilo\gram}$). (\textit{Left}) Picture of the room-temperature mockup rotating. The rotation is driven by a set of 64 coils mounted on the upper aluminum ring coupled with 8 small Neodymium motor magnets hosted in the rotating disk. The YBCO holder is removed to show the magnet ring (top right). (\textit{Right}) Detail of the Polyether ether ketone (PEEK) encoder holder coupled with 64 evenly spaced slits on the rotor.}

\end{figure}

The \textit{top panel} of Fig.~\ref{fig:mockup} shows a CAD cross section of the mockup: in the center there are 2 ball bearings (yellow) separated by an aluminum spacer. The umbrella support in the center positions a dummy HWP, which is composed of a lightened aluminum disk ($\sim$\SI{2}{\kilo\gram}) and the magnet ring (red). The stator is mounted in its final position and its YBCO bulks are in a normal state. The set of 64 coils (the smaller in blue) is positioned on the top of the external part of the main disk and is coupled with 8 small Neodymium motor magnets (\SI{8}{\milli\meter} diameter) on the rotating disk. On the same diameter there are 64 evenly spaced slits for the encoder readout system.
The \textit{left bottom panel} of Fig.~\ref{fig:mockup} shows the assembled system without the stator while \textit{right bottom panel} of Fig~\ref{fig:mockup} shows a of the Polyether ether ketone (PEEK) encoder holder which is coupled with 64 evenly spaced slits on the rotor.

\subsection{Motor driver and position readout electronics}
\label{sec:electronic}
In the prototype implementation, the coils mounted on the stator are powered in 8 groups. Together with the small magnets on the rotor, they form an 8-phase low-torque motor, optimized to minimize the heat losses in the system. A sampled, smoothed trapezoidal-wave, stored in permanent memory, is used to drive eight independent multiplying DACs and current generators. These produce 8 suitably phased currents, flowing in the 8 groups of coils. The phasing is such that when the current through a given coil is at the positive maximum, the current in the next coil is at the negative minimum, so that the first coil pushes the magnet while the next one pulls it. The rotation is sensed by an optical encoder, consisting of 64 precision machined, equally spaced slits, in the periphery of the rotor. Their position is read by means of LED emitters, optical fibers, and photodiodes, in the same way as in the Pilot experiment cryogenic WP rotator \cite{Salatino:article}. The measured rotation speed is compared to the desired rotation speed to produce an error signal, which is PID-processed and used to modulate the reference of the multiplying DACs, and thus the amplitude of the driving currents. For synchronization with the rest of the instrument, each transit of a slit below a fixed reference position is time-stamped with the value of a wide-counter, driven by the 5 MHz master clock of the instrument. An additional single slit, placed on a larger radius in the periphery of the rotor, is read in the same way. Its transit below the reference position resets a position counter, updated by the transit of each of the 64 slits. The position counter is then output together with its master clock time-stamp.

\subsection{Friction tests}
We first mounted only the bearing and the encoder in order to quantify the friction of the bearing with and without the driver motor magnets. The friction is quantified in terms of power loss and measured by spinning the rotor up to $\sim\SI{1.6}{\hertz}$ and then letting it free to slow down, while reading its angular position versus time with the optical encoder.
The rotation of the system is described by the equation of motion:
\begin{equation}
\label{eq:motion}
\tau(i) - \tau_f(\omega) = I\frac{d\omega}{dt},
\end{equation}
where $\tau$ is the external torque applied to spin the rotor, $\tau_f$ is the torque of friction forces, $I$ is the moment of inertia of the rotating system and $\omega$ the angular velocity of the rotor we measure. When the bearing is free to slow down (applied torque $\tau_a = 0$) we can convert Eq.~\ref{eq:motion} into an equation for the dissipated power:
\begin{equation}
\label{eq:torque_SMB}
\tau_f(\omega) = \frac{P_f(\omega)}{\omega} \, \, \rightarrow \, \, P_f(\omega) = -\omega I \frac{d\omega}{dt}.
\end{equation}

\begin{figure} [ht]
   \begin{center}
   \begin{tabular}{c} 
   \includegraphics[height=11cm]{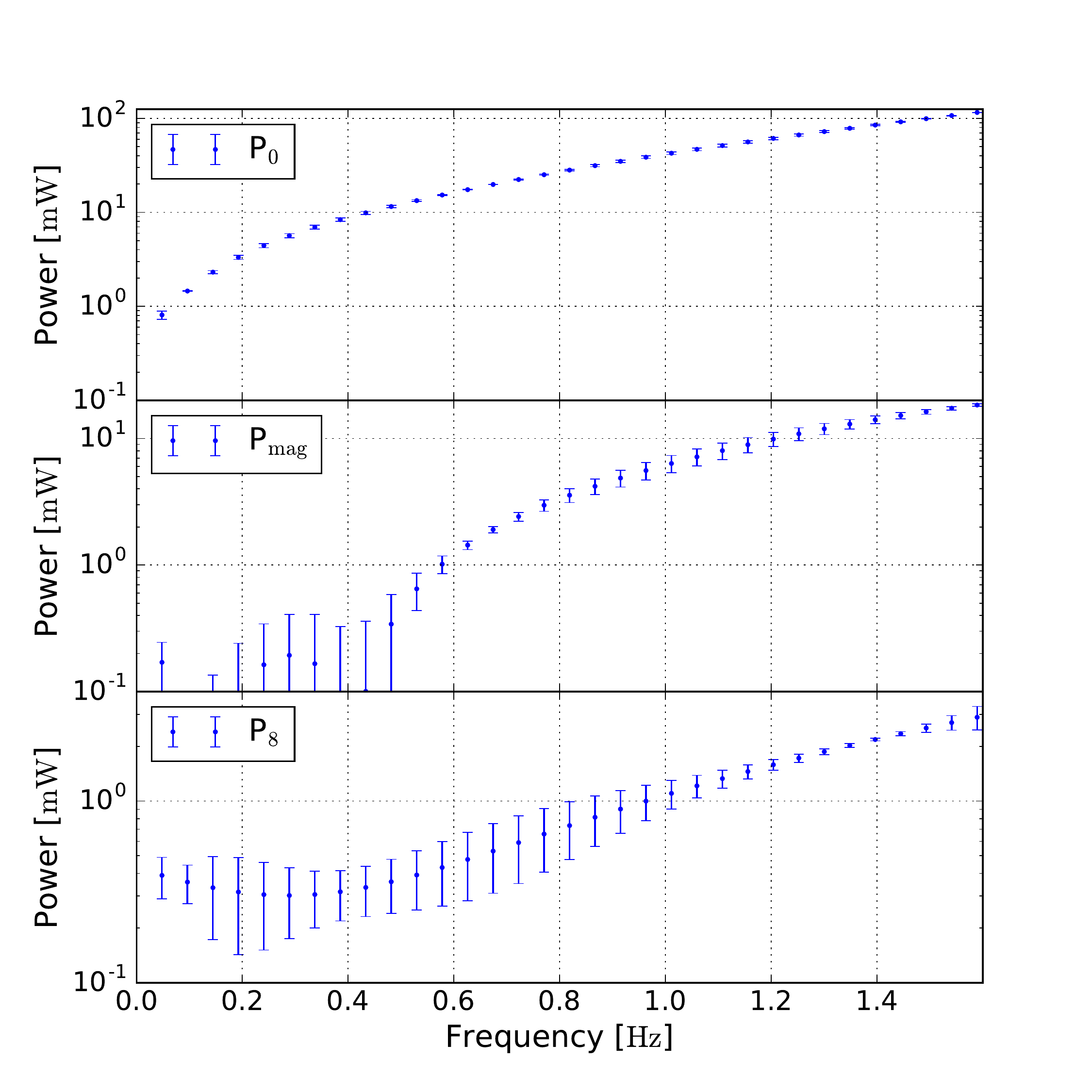}
   \end{tabular}
   \end{center}
   \caption[example] 
   { \label{fig:power_loss}
   Undersampled data of spin down tests calculated according to Eq.~\ref{eq:torque_SMB}. The power loss from friction is measured as a function of frequency for different contributions: ball bearings (${\rm P}_0$ in the \textit{top panel}), main magnet (${\rm P}_{\rm mag}$ in the \textit{central panel}) and 8 motor magnets (${\rm P}_8$ in the \textit{bottom panel:}). The main magnet and motor magnets contributions are differential measurements from the ball bearings contribution.
}
\end{figure}

Fig.~\ref{fig:power_loss} shows the spin down test performed for different configurations of the system in order to quantify the magnitude of each contribution to the total power budget.
The first configuration we tested consists only in the rotor. All conductive materials and the motor magnets are removed in order to measure the friction produced by the bearings (${\rm P}_0$). Than the whole system was assembled except for the motor magnets. This configuration allows us to quantify the eddy currents produced by the inhomogeneities ($\sim3\%$) of the main magnet (${\rm P}_{mag}$)\cite{Columbro:2021}. This value sets only an upper limit for the eddy currents in the cryo environment because the most of this contribution comes from the aluminum holder of the YBCO which will be mostly shielded by the superconductors at cryo temperature.
At the end we add the 8 motor magnets to determine their losses (${\rm P}_8$).

\subsection{Angular accuracy}
\begin{figure} [ht]
   \begin{center}
   \begin{tabular}{c} 
   \includegraphics[height=6cm]{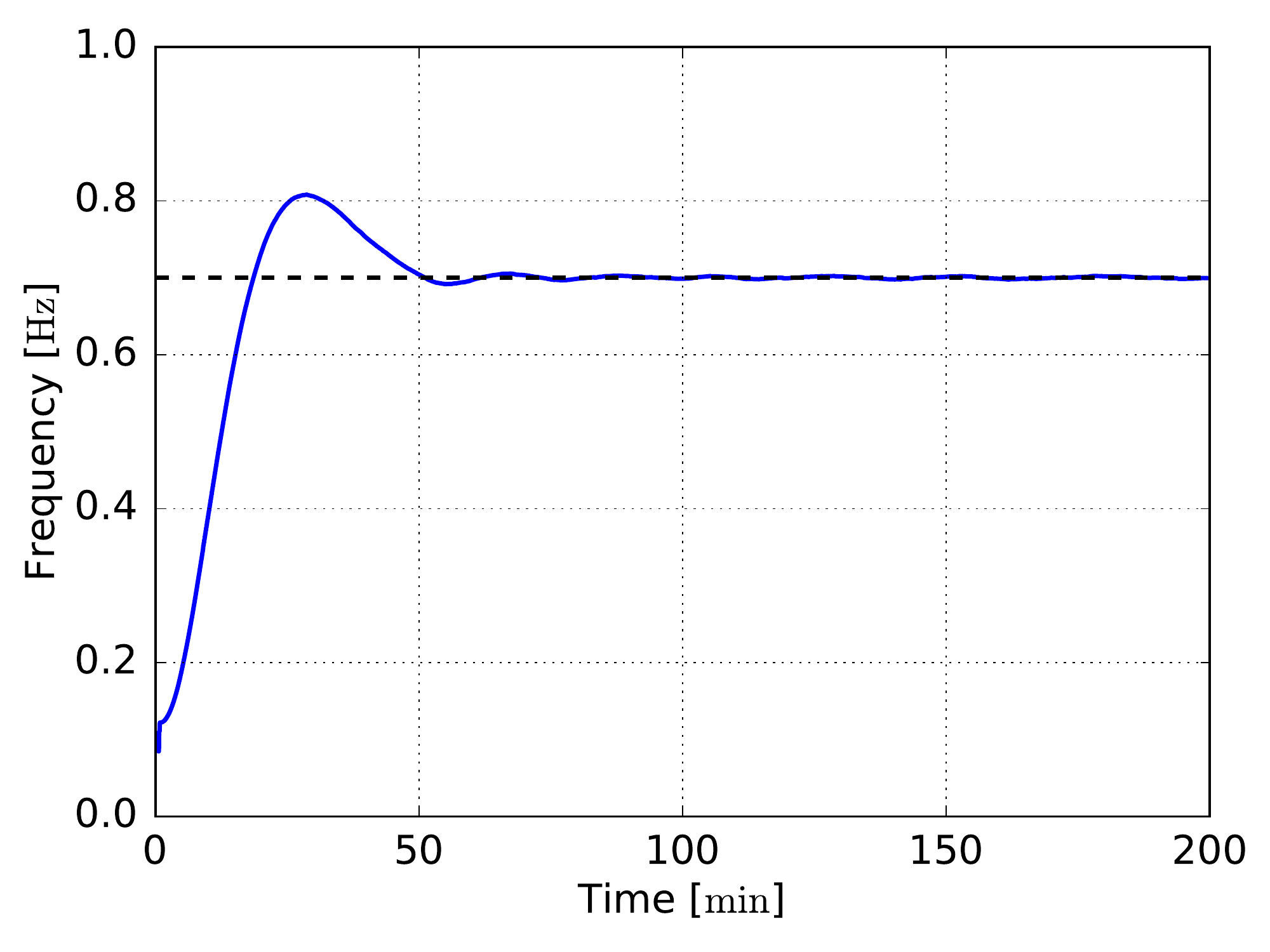}
\includegraphics[height=6cm]{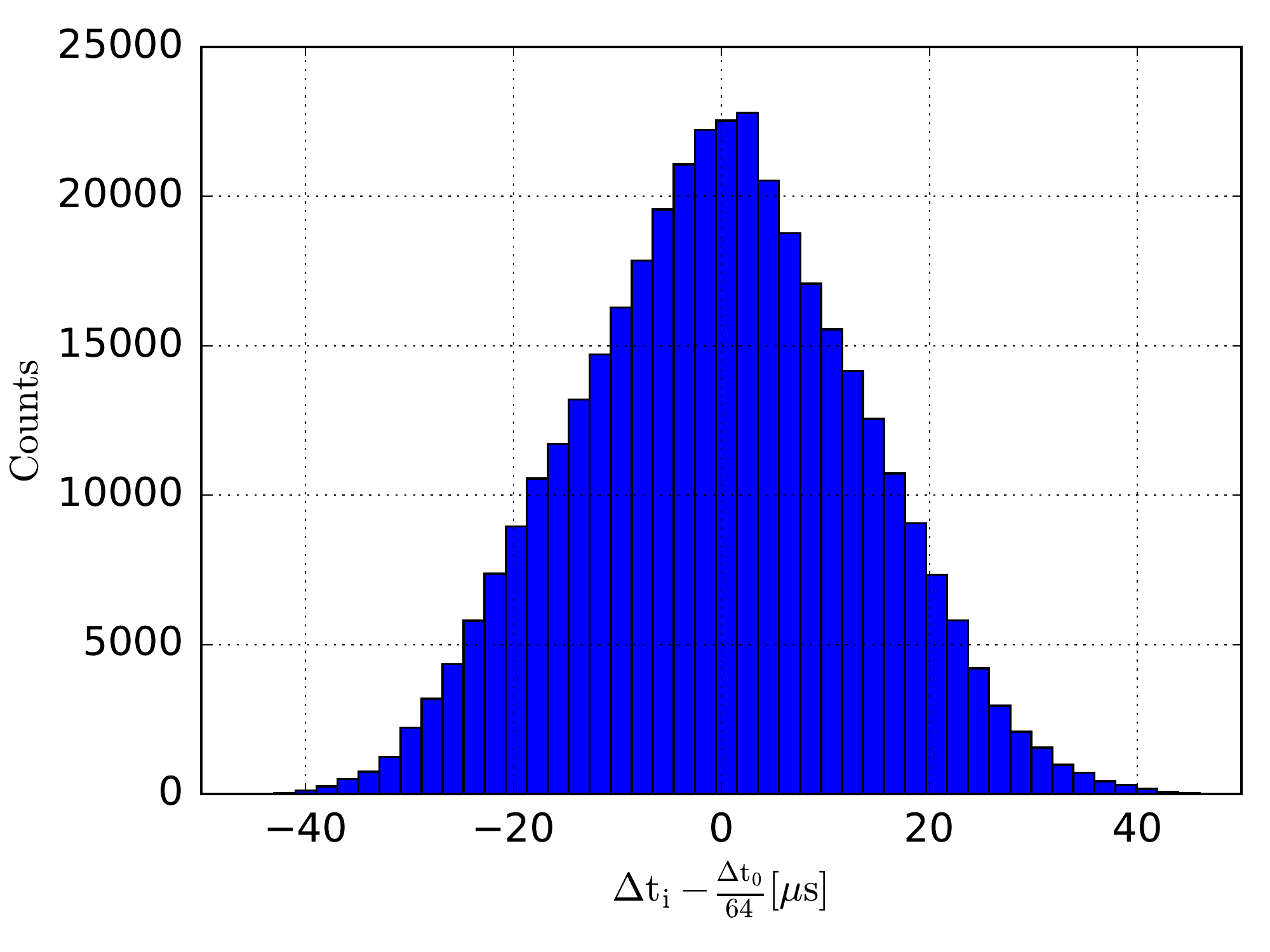}
   \end{tabular}
   \end{center}
   \caption[example] 
   { \label{fig:rotation}
     \textit{Left panel:} Relative encoder data acquired with a target frequency of \SI{0.7}{\hertz}. \textit{Right panel:} Histogram of the data during the stable rotation period ($\sim \SI{100}{\minute}$, $\sim$ 4200 rotations). 
   }
\end{figure}

The Proportional-Integrated-Derivative (PID) feedback controls both the frequency of pulses (allowing to spin up the rotor) and the magnitude of the current 32 times per round, to stabilize the rotation when the right frequency is reached. 
The user specifies the target frequency of the rotor which should be changed during the operation. Knowing the position of the 8 magnets (one every 8 slits), the relative phase of current in each series of coils is determined. The maximum value of the current is reached when the magnets are in the middle of two coils. Due to the inertia of the system, an additional small phase that is dependent on frequency is inserted to optimize system performance. Fig.~\ref{fig:rotation} shows a sample test performed with a target frequency of \SI{0.7}{\hertz} 

The PMU must be capable of reconstructing the HWP angle with high fidelity. Encoder performance at room temperature are fully representative of the cryogenic performance of the angular encoder system because it depends on properties which do not change with the temperature, including inertia of the system, warm readout electronics, rotational stability.
A rough estimate of the achievable angular accuracy in arcminute is calculated with the following relation:
\begin{equation}
 \sigma_\theta[\SI{}{\arcmin}] = \bar{\sigma} \cdot 360 \cdot f \cdot 60,
\end{equation}
where $f$ is the mean frequency expressed in \SI{}{\hertz} and $\bar{\sigma}$ is the mean value expressed in seconds of the standard deviations of the Gaussian distribution for each interval $\Delta t_i - \frac{\Delta t_0}{64}$, while $\Delta t_i$ and $\Delta t_0$ are the time in second reads 64 times per round by the relative encoder and only one time per round by the absolute encoder, respectively. The \textit{right panel} of Fig.~\ref{fig:rotation} shows the histogram of the data taken in the second half of the test shown in the \textit{left panel} of Fig.~\ref{fig:rotation}.

\begin{figure} [ht]
   \begin{center}
   \begin{tabular}{c} 
   \includegraphics[height=7cm]{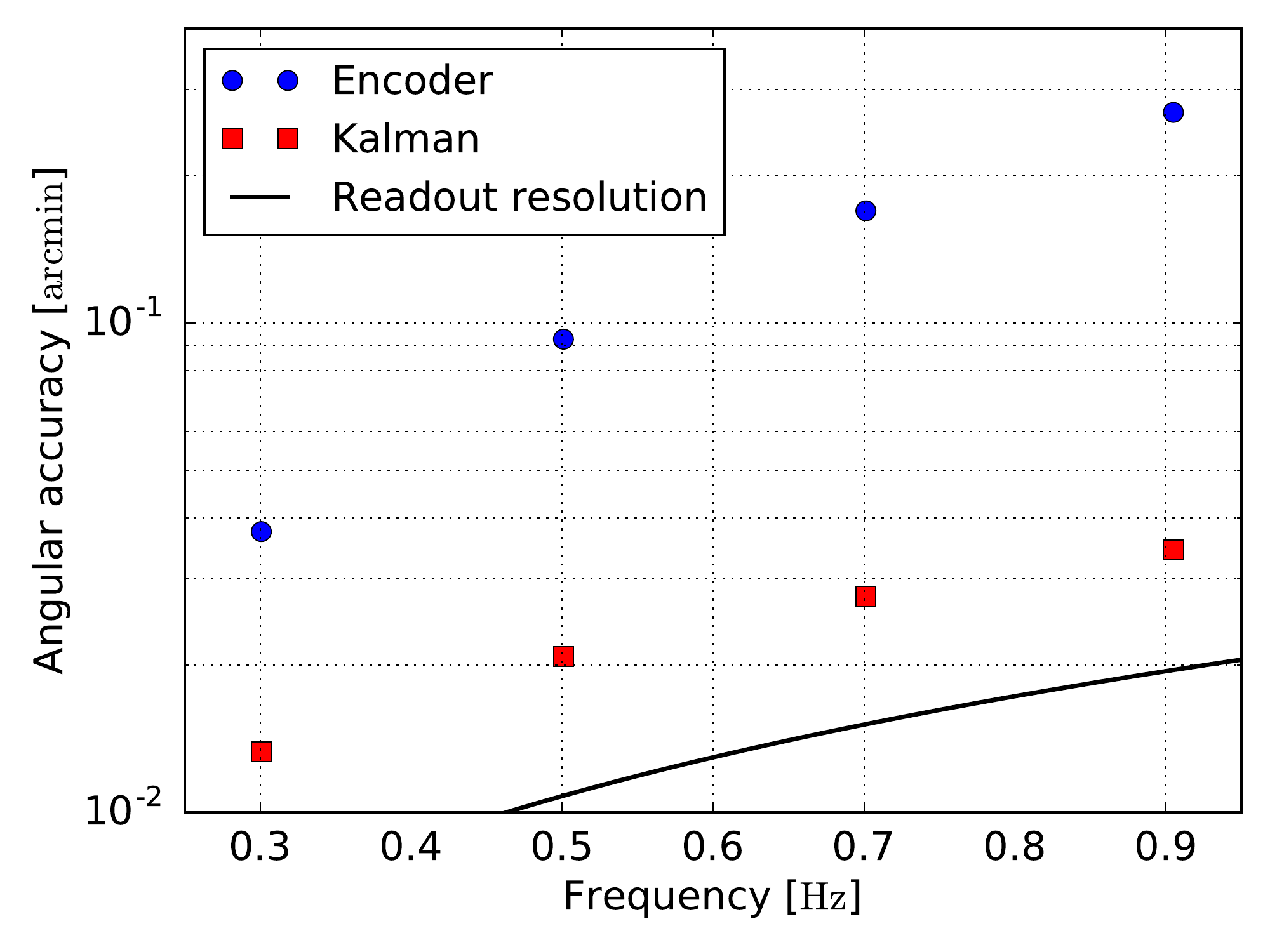}
   \end{tabular}
   \end{center}
   \caption[example] 
   { \label{fig:kalman}
The raw accuracy of the encoder data (blue dots), the accuracy resulting from the use of a Kalman filter  (red squares) and the accuracy corresponding to the electronic readout resolution (\SI{1}{\micro\second}).
   }
\end{figure}

Due to the high inertia of the system, all measurements of the position are correlated. We introduce a Kalman filter, which uses the dynamic model, the physical properties of the system, and multiple sequential measurements to make an estimate of the varying quantities that is better than the estimate obtained by using only one measurement alone.
The input parameters of the filter are the error on the readout data retrieved with the previous estimation and the uncertainty on the acceleration of the system estimated using the rotor inertia and the variation of the coil torque.
The accuracy improvement obtained by means of the Kalman filter ranges from a factor 3 at \SI{0.3}{\hertz} to a factor 8 at \SI{0.9}{\hertz}. The results of the Kalman filter application are shown in Fig.~\ref{fig:kalman} and are compared with the accuracy corresponding to the electronic readout resolution (\SI{1}{\micro\second}).
The current configuration is not limited by the electronic readout resolution but only by the stability of the rotation. This stability is limited by the current generator resolution which uses a 12-bit DAC. An improvement (from 12-bit to 16-bit) of the resolution is already planned but is not required for the SWIPE/LSPE modulator which has a requirement of \SI{0.05}{\arcmin} at \SI{0.5}{\hertz}.

\newpage

\section{Expected performance}
\label{sec:performance}
\subsection{Losses}
\label{sec:losses}
The tests performed with the LSPE/SWIPE prototype show that when the ball bearing friction is removed, the most important friction contribution comes from the inhomogeneities of the main magnet. The uniformity of the main magnet magnetic field should be improved by the help of the manufacturer up to 1\%. This is not so easy but seems feasible because of the smaller dimension of the permanent magnets with respect to the LSPE/SWIPE one\cite{Columbro:2021}. A further solution for HFT (due to the small radius of HFT magnet $\sim\SI{200}{\milli\meter}$) consists in the use of a single magnet with only one magnetization which will guarantee more uniformity.
By assuming the RRR = 2.8 for aluminum 6061-T6\cite{Duthil:article}, the dependence on frequency and magnetic dipole\cite{Reitz:article}, we can estimate the expected power loss produced by the rotor eddy currents: \SI{1.10}{\milli\watt} for MFT and \SI{1.45}{\milli\watt} for HFT\footnote{Starting from Fig.~\ref{fig:power_loss}, the power loss was estimated at the operating frequencies for both telescopes and by scaling the diameter values. For example the HFT (equivalent radius \SI{150}{\milli\watt}) eddy currents were estimated as:
\begin{equation}
P_8^\textit{HFT} = P_8(\SI{1.02}{\hertz}) \times \frac{d_\textit{HFT}^2}{d_\textit{mock}^2} \times RRR = \SI{1.25}{\milli\watt} \times \frac{150^2}{300^2} \times 2.8 = \SI{0.88}{\milli\watt}  .
\end{equation}
}

We expect hysteresis losses to be very small. This is due both to the absence of gravity which keeps in position the rotor after the release and to the high homogeneity of the magnetic field which minimizes hysteresis in the superconductor. We estimate\cite{Davis:article} a contribution $\ll\SI{0.5}{\milli\watt}$ which needs to be confirmed during cryogenic tests.

From parameters reported in Tab.~\ref{tab:MHFT_coils}, the resulting mean force for MFT (HFT) 8-phase motor is \SI{280}{\milli\newton\per\ampere} (\SI{414}{\milli\newton\per\ampere}). 
Assuming the same radius $R_*$ for all drag forces, we can find a rough estimate for the required force to spin the rotor:
\begin{equation}
F_{\textit{drag}} = \frac{P}{v} = \frac{P}{2\pi f R_{*}},
\end{equation}
which gives a required force of \SI{1.98}{\milli\newton} (\SI{2.26}{\milli\newton}), meaning that the current required is $\sim$\SI{7}{\milli\ampere} ($\sim$\SI{5}{\milli\ampere}) and the Joule loss is \SI{0.09}{\milli\watt} (\SI{0.05}{\milli\watt}).

As for the harness we decide to use manganin wires for the sensors and CuBe (\SI{0.25}{\milli\meter} diameter) for the motor and actuator wires in order to minimize the total heat load produced by the harness (\SI{0.22}{\milli\watt} for each telescope).

\begin{table}[ht]
\caption{Contribution to the power budget. The total expected heat load is $<\SI{4.19}{\milli\watt}$.}
\label{tab:losses}
\begin{center}       
\begin{tabular}{|l|l|l|} 
\hline
\rule[-1ex]{0pt}{3.5ex}   & \bf{MFT} & \bf{HFT}   \\
\hline
\rule[-1ex]{0pt}{3.5ex}   & [\SI{}{\milli\watt}] & [\SI{}{\milli\watt}]   \\
\hline
\rule[-1ex]{0pt}{3.5ex}  \textbf{8 magnets} & \SI{0.59}{} & \SI{0.88}{}  \\
\hline
\rule[-1ex]{0pt}{3.5ex}  \textbf{Main magnet} & $<$\SI{0.41}{} & $<$\SI{0.57}{}  \\
\hline
\rule[-1ex]{0pt}{3.5ex} \textbf{Hysteresis}  & $<$\SI{0.50}{} & $<$\SI{0.50}{}  \\
\hline
\rule[-1ex]{0pt}{3.5ex}  \textbf{Joule} & \SI{0.09}{} & \SI{0.05}{}  \\
\hline
\rule[-1ex]{0pt}{3.5ex} \textbf{Harness}  & \SI{0.22}{} & \SI{0.22}{} \\
\hline
\rule[-1ex]{0pt}{3.5ex} \textbf{Rotor emission}  & \SI{0.09}{} & \SI{0.07}{}  \\
\hline
\rule[-1ex]{0pt}{3.5ex} \textbf{Total}  & $<\SI{1.90}{}$ & $<\SI{2.29}{}$  \\
\hline
\end{tabular}
\end{center}
\end{table}

Tab.~\ref{tab:losses} summarizes all contributions to the power budget. The total expected heat load is \SI{4.19}{\milli\watt} which is of the order of the total power budget for both PMUs (\SI{4}{\milli\watt}). Because this estimate was made by using the upper limit both of the main magnet and hysteresis contributions, there are margins to be within the budget. 

The main contribution to the motor eddy currents comes from the aluminum holder of the YBCO. Making the holder of electrical insulator like G10 will remove eddy currents but will thermally insulate the superconductor ring. This may make the cooldown time of the superconductor too long. The possibility of using an electrical insulator for the upper part of the holder and a thermal conductor (aluminum) for the lower part is under study.

\subsection{HWP Temperature}
\begin{figure} [ht]
   \begin{center}
   \begin{tabular}{c} 
   \includegraphics[height=7cm]{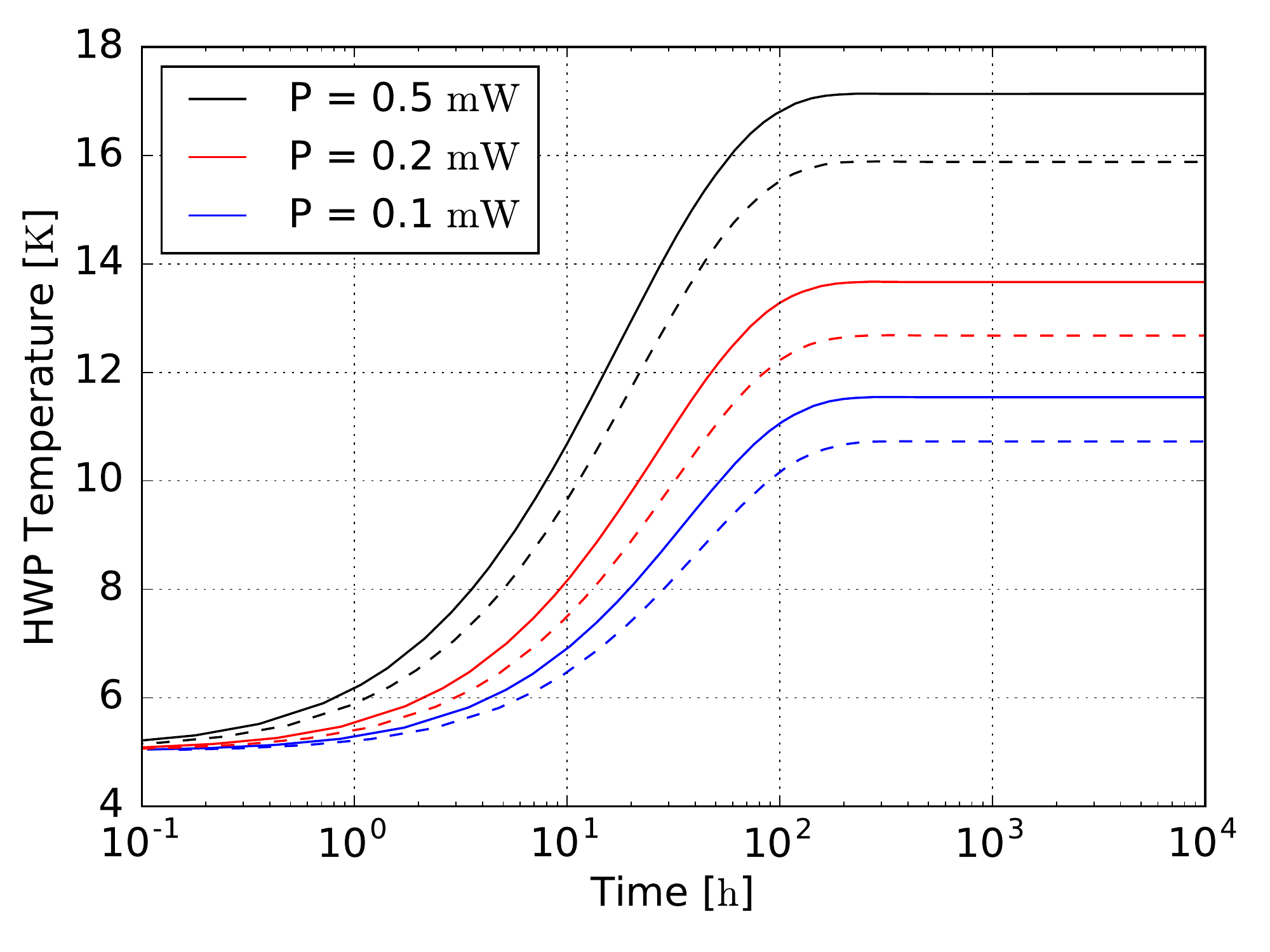}
   \end{tabular}
   \end{center}
   \caption[example] 
   { \label{fig:thermal_model}
Temperature of the HWPs as a function of time for different values of power load on the rotor (\SI{0.1}{\milli\watt} is the most likely case). Dashed lines correspond to the MFT and solid lines to the HFT.
   }
\end{figure}

The temperature of both HWPs must be $< \SI{20}{\kelvin}$ to reduce the radiative loading on the detector and minimize the amplitude of spurious signals\cite{Columbro_systematics:article, Hileman:article, Salatino:article}.
We use Comsol Multiphysics to build a thermal model of the rotor surrounded by a \SI{5}{\kelvin} environment. The rotor is made by Aluminum for the encoder and the groove rings, while the material assumed for the magnetic ring is the iron. This choice is driven by the lack of SmCo measured properties at very low temperature (iron physical properties are very similar to SmCo up to \SI{77}{kelvin}). The estimated main band HWP emissivities are \SI{0.02}{} and \SI{0.03}{} for MFT and HFT (see Fig.~\ref{fig:hwp_band}), respectively, while the emissivity outside the instrument bands is \SI{0.03}{}. The assumed emissivity of aluminum is \SI{0.5}{} when the surface is blackened. The expected dissipation on the rotor is $\sim\SI{0.1}{\milli\watt}$ which mainly comes from the modulated current in the motor coils. 
The heating propagated on the edge of the encoder ring (where the coils are located) are analyzed also for more pessimistic cases (\SI{0.2}{\milli\watt} and \SI{0.5}{\milli\watt}).
The real HWP is made of polypropylene and Cu meshes: the polypropylene has strong absorption features in the thermal IR (which help to cool the HWP more quickly in the initial stages) and high transparency at long wavelengths while the inductive Cu meshes has high emissivity at low frequency. In these frequency range the most relevant heat sources for the HWP are interplanetary dust (IPD) emission and instrument emission.
While the instrument emission is taken into account in the simulation with a \SI{5}{\kelvin} environment, the IPD is not and varies across the sky with a smooth distribution. This results in a sine-like time profile with the same period as the satellite spin.
The total radiative load at all ecliptic latitudes is of the order of a few \SI{}{\micro\watt}, which is negligible with respect to the heating from the coils. Combined with a HWP thermal time constant of about \SI{10}{}-\SI{20}{\hour} (from simulation of Fig.~\ref{fig:thermal_model}) this produces a negligible sky-synchronous variation of the HWP temperature, resulting in a negligible loading variation on the detectors.

Fig.~\ref{fig:thermal_model} shows that the rotor (HFT solid lines, MFT dashed lines) reaches the equilibrium temperature of $<\SI{20}{\kelvin}$ within a few days under all scenarios that were modeled, minimizing the impact on the detector background and instrument sensitivity.

\subsection{Angular accuracy}
The LiteBIRD total angular error budget corresponds to \SI{1}{\arcmin} for MFT and \SI{5}{\arcmin} for HFT and is equally split in 3 error contributions: angle reconstruction, positioning of the HWP reference, margin.

As stated before, the angular accuracy of the system is related to the inertia of the rotor, to its speed and to the warm readout resolution. The main parameters and expected performance of LiteBIRD PMUs are summarized in Tab.~\ref{tab:accuracy}.
The MFT modulator is very similar to the prototype tested configuration and a similar performance is expected. The faster rotation and lower inertia of the HFT modulator reduce rotational stabilization, resulting in a reduced angular accuracy. All the same, the HFT raw encoder accuracy is nearly sufficient to meet the requirement and can be readily improved to perform well below the requirement by use of a Kalman filter. The improvement can be achieved by increasing the current generator resolution to have a finer control of the motor current in the PID feedback loop.

\begin{table}[ht]
\caption{Main parameters of LSPE/SWIPE, MFT and HFT configurations. The encoding accuracy of LiteBIRD modulators is estimated using the same configuration used in LSPE/SWIPE.}
\label{tab:accuracy}
\begin{center}       
\begin{tabular}{|l|l|l|l|l|} 
\hline
\rule[-1ex]{0pt}{3.5ex}   &  & \bf{SWIPE} & \bf{MFT} & \bf{HFT}   \\
\hline
\rule[-1ex]{0pt}{3.5ex}  \bf{HWP diameter} & \SI{}{\milli\meter} & \SI{500}{} & \SI{320}{} & \SI{220}{}  \\
\hline
\rule[-1ex]{0pt}{3.5ex}  \bf{Frequency} & rpm & \SI{30}{} & \SI{39}{} & \SI{71}{} \\
\hline 
\rule[-1ex]{0pt}{3.5ex}  \bf{Encoder speed} & \SI{}{\meter\per\second} & \SI{1.0}{} & \SI{1.0}{} & \SI{1.3}{}  \\
\hline 
\rule[-1ex]{0pt}{3.5ex}  \bf{Moment of inertia} & \SI{}{\kilogram\meter^2} & \SI{0.5}{} & $\sim\SI{0.2}{}$ & $\sim\SI{0.05}{}$  \\
\hline 
\rule[-1ex]{0pt}{3.5ex}  \bf{Encoding accuracy} & \SI{}{\arcmin} & \SI{0.09}{} & $\sim\SI{0.4}{}$ & $\sim\SI{5.7}{}$  \\
\hline 
\rule[-1ex]{0pt}{3.5ex}  \bf{Kalman accuracy} & \SI{}{\arcmin} & \SI{0.02}{} & $\sim\SI{0.1}{}$ & $\sim\SI{0.7}{}$  \\
\hline 
\end{tabular}
\end{center}
\end{table}

\section{Conclusions}
We presented the baseline design of LiteBIRD PMUs for the mid and high frequency telescopes. Both PMU are located at \SI{5}{\kelvin} and based on a continuously transmissive rotating HWP which has a transmission across the bands on average greater than 95\%.
We discussed the tests performed on a room-temperature rotating disk used as stand-in for the cryogenic HWP rotor which helped in the confirmation of the models used for the LiteBIRD design. The expected total load on the \SI{5}{\kelvin} stage is $<\SI{4.19}{\milli\watt}$ which is close to the requirement of \SI{4}{\milli\watt}. The angular accuracy in the angle reconstruction is $\SI{0.4}{\arcmin}$ (\SI{5.7}{\arcmin}) for MFT (HFT). The introduction of a Kalman filter improves the accuracy of angle reconstruction down to \SI{0.1}{\arcmin} and \SI{0.7}{\arcmin} for MFT and HFT, both values lower than the requirements of \SI{1}{\arcmin} and \SI{5}{\arcmin}, respectively.
Both HWP temperatures are expected to be below \SI{20}{\kelvin}, the maximum value to minimize the impact on the detector background and on instrument sensitivity.
In conclusion, all values are close to the requirement. In any case, they represent the worst case for the expected performance, giving us some design margin.

\newpage

\acknowledgments 

This work is supported in \textbf{Japan} by ISAS/JAXA for Pre-Phase A2 studies, by the acceleration program of JAXA research and development directorate, by the World Premier International Research Center Initiative (WPI) of MEXT, by the JSPS Core-to-Core Program of A. Advanced Research Networks, and by JSPS KAKENHI Grant Numbers JP15H05891, JP17H01115, and JP17H01125. The \textbf{Italian} LiteBIRD phase A contribution is supported by the Italian Space Agency (ASI Grants No. 2020-9-HH.0 and 2016-24-H.1-2018), the National Institute for Nuclear Physics (INFN) and the National Institute for Astrophysics (INAF). The \textbf{French} LiteBIRD phase A contribution is supported by the Centre National d’Etudes Spatiale (CNES), by the Centre National de la Recherche Scientifique (CNRS), and by the Commissariat à l’Energie Atomique (CEA). The \textbf{Canadian} contribution is supported by the Canadian Space Agency. The \textbf{US} contribution is supported by NASA grant no. 80NSSC18K0132. 
\textbf{Norwegian} participation in LiteBIRD is supported by the Research Council of Norway (Grant No. 263011). The \textbf{Spanish} LiteBIRD phase A contribution is supported by the Spanish Agencia Estatal de Investigación (AEI), project refs. PID2019-110610RB-C21 and AYA2017-84185-P. Funds that support the \textbf{Swedish} contributions come from the Swedish National Space Agency (SNSA/Rymdstyrelsen) and the Swedish Research Council (Reg. no. 2019-03959). The \textbf{German} participation in LiteBIRD is supported in part by the Excellence Cluster ORIGINS, which is funded by the Deutsche Forschungsgemeinschaft (DFG, German Research Foundation) under Germany’s Excellence Strategy (Grant No. EXC-2094 - 390783311). This research used resources of the Central Computing System owned and operated by the Computing Research Center at KEK, as well as resources of the National Energy Research Scientific Computing Center, a DOE Office of Science User Facility supported by the Office of Science of the U.S. Department of Energy.

\bibliography{report} 
\bibliographystyle{spiebib} 

\end{document}